\documentclass[10pt]{article}
\usepackage{amssymb}
\usepackage{amsthm}
\usepackage[pdftex]{graphicx}

\newcommand{\cm}{cm$^{-1}$}
\newcommand{\ccm}{cm$^{-3}$}
\newcommand{\q}{$|q|$}
\begin{document}
\begin{center}
{\Large \textbf{Asymmetry to symmetry transition of Fano line-shape: Analytical description}}

\vspace{0.25cm}
{\large Rajesh Kumar} \footnote{Corresponding authors: rajeshkumar@iiti.ac.in}

\vspace{0.25cm}
Department of Physics, School of Basic Sciences, Indian Institute of Technology Indore, IET-DAVV Campus, Indore, Madgya Pradesh 452017, India
\end{center}

\begin{abstract}
An analytical description of Fano line-shape asymmetry ratio has
been presented here for a general case. It is shown that Fano
line-shape becomes less asymmetric as \q is increased and finally
becomes completely symmetric in the limiting condition of q equal to
infinity. Asymmetry ratios of Fano line-shapes have been calculated
and are found to be in good consonance with the reported expressions
for asymmetry ratio as a function of Fano parameter. Application of
this derivation is also mentioned for explanation of asymmetry to
symmetry transition of Fano line-shape in quantum confined silicon
nanostructures.

\textbf{Keywords:}  Electron-phonon interaction, Raman lineshape,
Fano line-shape.

\textbf{PACS No.:}  78.67.f; 63.22.-m; 78.30.j

\end{abstract}

\section{Introduction}

Fano resonance  \cite{1} is a general property of the physical
system where a degeneracy of continuum states and a discrete level
raised into this continuum is present. Such a system is possible in
simple systems like atoms [2], solids  [3,4] as well as in
complicated semiconductor systems like quantum heterostructures [5].
Fano resonance, also referred as Fano effect, can be observed in two
different regimes namely linear and nonlinear. The linear regime is
old and well understood and the effect is explained by classical
theory given by Fano in his original paper  [1]. Fano effect got
importance after the observation of non-linear Fano effect very
recently  [6,7] in quantum systems. However the discussion will be
limited to the classical Fano effect in this paper. In solid state,
the Fano effect is understood in terms of electron-phonon
interaction  [8-10] provided the solid state system contains an
available discrete level and a filled continuum of states. Apart
from bulk materials, Fano effect has been observed in nanostructures
[11-13] and heavily doped semiconductors  [14-16]. In
nanostructures, discrete phonons can interfere with continuum of
electronic states available in the material as a result of quantum
confinement  [17-19]. On the other hand, in heavily doped
semiconductors, the continuum of electronic states is available due
to intra sub-band excitations possible due to Fermi level being
pushed as a result of ultra-heavy doping in degenerate semiconductor
[4,10,20,21].  Irrespective of the exact process, the result of this
interaction is a characteristic asymmetric line-shape also known as
Fano line-shape. Different properties of this line-shape is
investigated to understand different physical systems. Understanding
of Fano effect is important in exploring different fields of
electronic transport  [20,22] and spectroscopy  [5,23].
Understanding this effect also helps in understanding different
phenomena happening at microscopic level within the system  [24].
One of the properties of this line shape is its asymmetry and the
condition under which it becomes symmetry. This aspect will be dealt
in this paper.

U. Fano in his classic observed this type of asymmetry in atomic
systems  [1]. He also indicated the possibility of this type of
interaction in quantized discrete and continuous energy states. In
other words, the Fano interaction can only be observed in systems
where a discrete state is located within the continuum of different
type of energy states. Semiconductors can be this type of classic
system, where phonons have discrete energy states. The continuum of
states may be provided by the electronic states under special
conditions like heavy doping as in the case of degenerate
semiconductors. Different conditions required for this type of
interaction will be discussed in details in the following
subsections. The possible Fano interaction between continuum of
electronic states and discrete phonon state can be studied by Raman
scattering  [25,26] or IR spectroscopy [27]. The general properties
of Fano effect can be understood by considering the special case of
heavily doped Si  [3,4,10,28,29] and ion-implanted Si [30], where
Fano interaction is observed if the level of doping is of the order
of $10^{19}$ \ccm, the semiconductor behaves like a degenerate
semiconductor. Position of the Fermi level goes deep inside the
conduction band, which gives rise to continuous electronic Raman
scattering between the $\Delta_1$  and  $\Delta_2$ bands along (100)
directions [3,4]. A continuum of electronic transitions exists from
a minimum energy ($\hbar \omega_{min}$) [10] to a maximum energy
($\hbar \omega_{max}$). These extreme energies depend on the
position of the Fermi level in the conduction band. In such systems,
interference between a discrete phonon state and continuum of
electronic states take place if the discrete optical phonons with
energy $\hbar \omega_{0}$ satisfies the following condition given by
eq. 1:
\begin{equation}\label{condition}
     \hbar \omega_{min} < \hbar \omega_{0} <\hbar \omega_{max}
\end{equation}

A system that satisfies the condition given in eq. \ref {condition},
can show Fano-type asymmetric line-shape. If Raman spectroscopy is
used to characterize the Fano effect, the generalized Raman
line-shape as a result of Fano interaction can be written as
following Eq. 2:
\begin{equation}\label{fano}
    I_F (\omega) = \frac{(q+\varepsilon)^2}{1+\varepsilon^2}
\end{equation}

where $\varepsilon = \frac{\omega - \omega _0}{\gamma/2}$ with
$\gamma$ as the full width at half maxima (FWHM) and $\omega _0$ is
the observed wavenumber of the Fano transition. The `\q' is the Fano
asymmetry parameter, which provides the measure of Fano
interference. A low value of `\q' means strong electron-phonon
interaction. Asymmetric Raman line-shape is a signature of
electron-phonon (Fano) interaction. Fano-type asymmetry in the Raman
line-shape can be diagnosed by the following properties, which holds
true for Fano line-shape. Upon varying the exciting laser
wavelength, Raman peak position is unchanged whereas asymmetry
parameter (\q) increases with decreasing excitation wavelengh
[10,31]. The interplay between \q value and the asymmetry of the
line-shape is extremely important to understand a system showing
Fano resonance. Aim of this paper is to give an analytic derivation
to prove the asymmetry to symmetry conversion of Fano line-shape
function.

\section{Analysis and discussion}
The Fano function given by Eq. 2 can be symmetric or asymmetric
depending on the magnitude of \q  with a maximum value at
    \begin{equation}\label{3}
    \omega_{maximum} = \omega_0 + \frac{\gamma}{2q}
\end{equation}

As evident from this equation, Fano asymmetry parameter, \q,
determines the peak position and asymmetry of the line-shape.
Quantitatively asymmetry ratio is defined here as $\gamma_a
/\gamma_b$, where $\gamma_a$ and $\gamma_b$ are half widths on the
low- and high-energy side of the maximum. The magnitude of \q
decides the value of asymmetry whereas sign decides whether
asymmetry is in higher wavenumber side or the lower side. In this
paper, only positive values of \q is discussed in which case the
Fano line-shape is wider in the higher wavenumber side. The Fano
line-shape becomes symmetric for \q $\sim \infty$. This property of
Fano line-shape can be seen experimentally as well as numerically.

An analytic derivation for asymmetric to symmetric transition of
line-shape given in Eq. 2 will be derived here to understand Fano
effect in a better way. Figure 1 Shows a typical line-shape given by
Eq. 2 for different values of \q. the value of $\omega_0$  and
$\gamma$ were taken as 520 \cm and 8\cm respectively. For \q
$=\infty$, the curve is symmetric with peak at 520 \cm as shown in
figure 1(a). On the other hand, the Fano line-shape gets inverted by
keeping the symmetry intact and shows a dip at 520 \cm in stead of a
peak as can be seen in figure 1(b). For intermediate value of \q =1,
a typical Fano line shape can be seen in figure.1(c). However,
figure 1(c) shows a maximum at 524 \cm (value is consistent with eq.
3) asymmetry ratio can't be defined for the same. On the other hand,
the observation of minimum point at 516 \cm is typical observation
when \q $\sim$ 1. Observation of such minima in this regime is
called the antiresonance [4] and is a characteristic of high Fano
coupling.

Figure 2(a) to 2(f) shows the line-shape given in Eq. 2 for
different \q values of 2.5, 3, 4, 5, 7.5 and 10 respectively. A \q
dependent asymmetry and antiresonance can be seen in figure 2
clearly. As the value of \q is decreasing (a to f), line-shape
becomes more and more asymmetric with the curve getting broader on
the higher wavenumber side. In other words, the Fano line-shape is
less asymmetric for higher values of \q. This relationship can be
derived as follows:

For, $\omega = \omega_0$, in Eq. 2, $\varepsilon = 0$ and $I_F$ has
a maximum value equal to $I_F^{max}(\omega) = q^2$. Since, $\gamma$
is the FWHM which means $\omega = \omega_0 + \frac{\gamma}{2}$ and
$\omega = \omega_0 - \frac{\gamma}{2}$ should be the points of equal
intensities for the case of a near symmetric line-shape.
Theoretically at $\omega = \omega_0 + \frac{\gamma}{2}, \varepsilon
= 1$ and
\begin{equation}\label{4}
    I_F=\frac{(q+1)^2}{2}=I_F^+ (Say)
\end{equation}

Similarly, at $\omega = \omega_0 - \frac{\gamma}{2}, \varepsilon =
-1$ and
\begin{equation}\label{5}
    I_F=\frac{(q-1)^2}{2}=I_F^- (Say)
\end{equation}

From Eqs. (4) and (5), it is clear that intensity at $\omega =
\omega_0 + \frac{\gamma}{2}$ and $\omega = \omega_0 -
\frac{\gamma}{2}$ not equal. It means that the line-shape given by
the Eq. (2) is asymmetric for \q$
>1$. Thus the ratio of intensities at $\omega =
\omega_0 + \frac{\gamma}{2}$ $\omega = \omega_0 - \frac{\gamma}{2}$
will be given by the Eq. (6) as follows.

\begin{equation}\label{6}
    \frac{I_F^+}{I_F^-}=\frac{(q+1)^2}{(q-1)^2} = \frac{(1+\frac{1}{q})^2}{(1-\frac{1}{q})^2}
\end{equation}

\begin{equation}\label{7}
    \lim_{q\rightarrow \infty} \frac{I_F^+}{I_F^-} = 1
\end{equation}

It is clear from Eq. (7) that for `\q' tending to infinity, the Eq.
(2) is symmetric in nature with FWHM of $\gamma$.    This analytic
derivation is consistent with theoretically plotted line-shapes.
This property has been used in silicon nanostructures to explain the
simultaneous observation of Fano effect and quantum confinement
effect. In the work reported by Kumar et al  [11], authors show that
only quantum confinement effect is observed under control excitation
condition to avoid any Fano effect. This observed fact is proved
here analytically.    Observed asymmetry ratio form line-shape in
fig2 have been plotted as a function of Fano asymmetry parameter in
figure 3 as solid points. The dotted line in figure 3 represents the
theoretical asymmetry ratio of line-shape in eq 2. The line in
figure 3 is expressed as eq. (8)  [28] as follows:
\begin{equation}\label{8}
    Asymmetry ratio (q)= \frac{q+1}{q-1}
\end{equation}

Very good match between theoretically predicted and numerically
calculated asymmetry ratios can be seen in figure 3. This match
further confirms the fact that Fano line-shape in eq. 2 converges to
symmetric one for low coupling regime characterized by higher values
of \q.

\section{Conclusion}
Theoretical analysis of Fano line-shape asymmetry ratio show that
Fano line-shape is asymmetric for Fano parameters (\q) $>$1. The
Fano-line shape is characterized by the point of antiresonance for
\q $\sim$ 1. Calculated Fano line-shape show that Fano line-shape
becomes less asymmetric as \q is increased and finally becomes
completely symmetric in the limiting condition of \q equal to
$\infty$. This derivation has been used to explain the experimental
observation of one of the old reports for the observation of
photoexcited Fano interaction in silicon nanostructures.

\newpage

\begin{figure}
\begin{center}
\includegraphics[width=15cm]{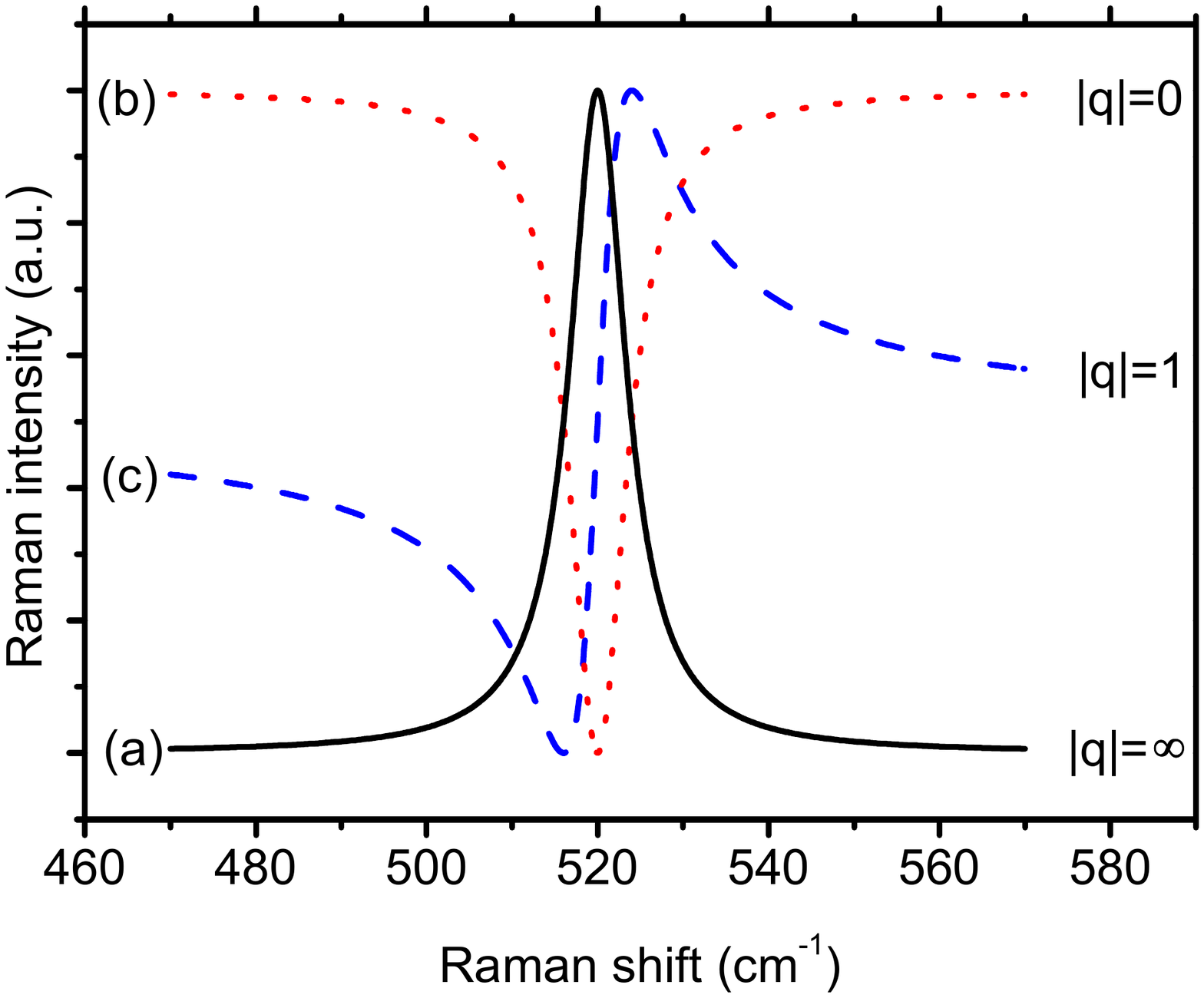}
\caption{Raman line-shape described by Eq. 2 for different values of
\q = (a) $\infty$, (b) 0 and (c) 1}
\end{center}
\end{figure}

\begin{figure}
\begin{center}
\includegraphics[width=15cm]{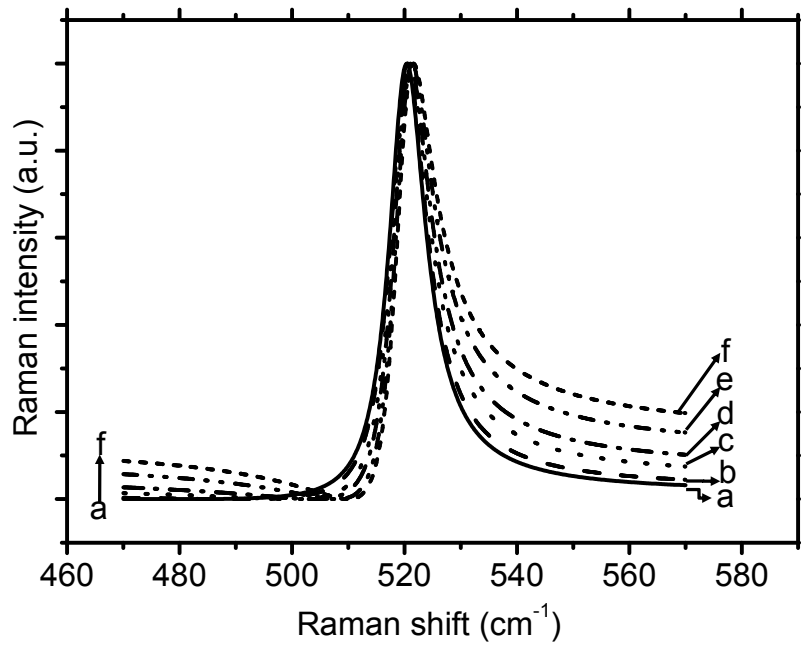}
\caption{Raman line-shape described by Eq. 2 for different values of
\q = (a) 10, (b) 7.5, (c) 5, (d) 4, (e) 3 and (f) 2.5}
\end{center}
\end{figure}

\begin{figure}
\begin{center}
\includegraphics[width=15cm]{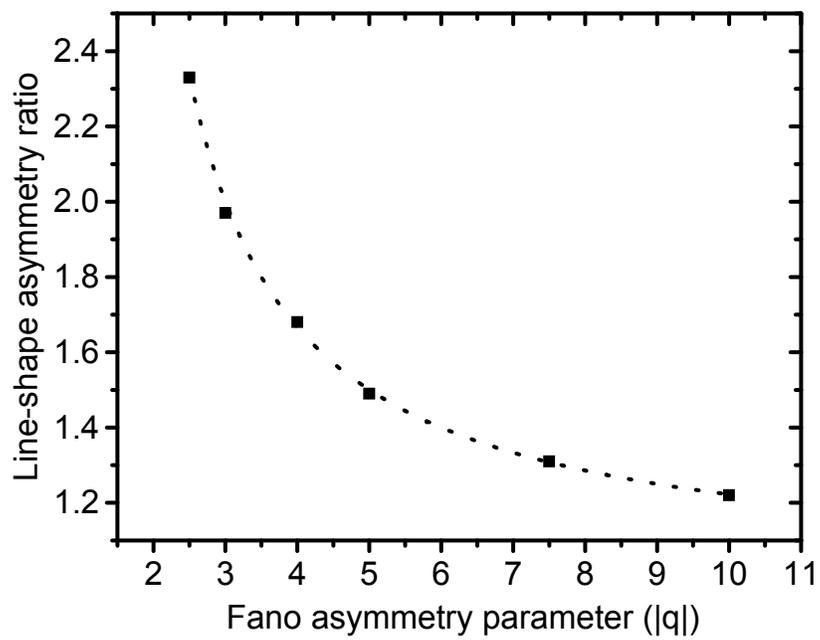}
\caption{Raman line-shape asymmetry ratio, obtained from Fig 2, as a
function of Fano parameter \q (points). Plot of Eq. 8 is shown as
dotted line.}
\end{center}
\end{figure}

\end{document}